\documentclass[12pt]{iopart}
\usepackage{graphicx}
\begin{document}

\title{Two-dimensional easy-plane SU$(3)$ magnet with the transverse field:
Anisotropy-driven multicriticality}

\author{Yoshihiro Nishiyama}

\address{Department of Physics, Faculty of Science,
Okayama University, Okayama 700-8530, Japan}
\vspace{10pt}

\begin{abstract}
The two-dimensional easy-plane
SU$(3)$ magnet subjected to the transverse field
was investigated with the exact-diagonalization method.
So far, as to the $XY$ model (namely, the easy-plane SU$(2)$ magnet),
the transverse-field-driven order-disorder phase boundary 
has been investigated with the exact-diagonalization method,
and it was claimed that 
the end-point singularity (multicriticality)
at 
the $XX$-symmetric point does not accord with large-$N$-theory's prediction.
Aiming to reconcile the discrepancy,
we extend the internal symmetry to the easy-plane SU$(3)$
with the anisotropy parameter $\eta$,
which interpolates
the isotropic ($\eta=0$)
and fully anisotropic ($\eta=1$) cases smoothly.
As a preliminary survey, setting $\eta=1$,
we analyze the order-disorder phase transition 
through resorting to
the fidelity susceptibility $\chi_F$, which
exhibits a pronounced signature for the criticality.
Thereby, with $\eta$ scaled carefully,
the $\chi_F$ data are cast into the crossover-scaling formula
so as to determine the crossover exponent $\phi$,
which seems to reflect the extension of the internal symmetry group
	to SU(3).

\end{abstract}

%
%
%
%
%

\section{\label{section1}Introduction}

The one-dimensional $XY$ model 
subjected to the transverse field $H$
and anisotropy $\eta$
with the Hamiltonian
${\cal H}_{XY} =-\sum_i [(1+\eta)\sigma^x_i\sigma^x_{i+1}
+(1-\eta)\sigma^y_i\sigma^y_{i+1}+H\sigma^z_i]$
(${\mathbf \sigma}_i$: Pauli matrices at site $i$)
is attracting much attention 
\cite{Maziero10,Sun14,Karpat14,Luo18}
in the context of the quantum information theory
\cite{Steane98,Bennett00}.
A key ingredient is that
the model covers  both $XX$-  ($\eta=0$)
and Ising-symmetric ($\eta=1$)
cases,
and there appear rich characters as to
the transverse-field-driven order-disorder phase transition.
Recently,
the multicriticality at $\eta=0$,
{\it i.e.}, the end-point singularity of the phase boundary 
toward the $XX$-symmetric point,
was explored in depth from the quantum-information-theoretical viewpoint
\cite{Mukherjee11}.

Meanwhile, its extention to the two-dimensional counterpart has been made
\cite{Henkel84,Wald15}.
According to the large-$N$ theory \cite{Wald15}, namely,
for sufficiently large internal symmetry group,
the phase boundary should exhibit a reentrant 
(non-monotonic) behavior.
The reentrant behavior leads to  a counterintuitive picture
such that the disorder phase is induced by the lower internal symmetry group.
On the contrary, as to the $XY$ (namely, easy-plane SU(2)) model,
the exact-diagonalization study \cite{Henkel84,Wald15,Nishiyama19}
claimed that
the phase boundary rises up linearly (monotonically) in proximity to the multicritical point;
the similar phase diagram was reported in Ref. \cite{Jalal16},
where a topological index is specified to each phase surrounding the multicritical point.
Rather technically, it has to be stressed that
the exact-diagonalization method
allows direct 
access to the ground state (namely, infinite imaginary-time system size),
and hence, we do not have to care about
the anisotropy between the real-space and imaginary-time directions
rendered by the
dynamical critical exponent
$z=2(\ne 1)$ \cite{Zapf14} at the multicritical point.

The aim of this paper is to 
reconcile the discrepancy
between the large-$N$- 
\cite{Wald15}
and easy-plane-SU$(2)$-based 
\cite{Henkel84,Wald15,Nishiyama19}
results;
namely, so far, only the extremum cases have been considered.
For that purpose,
we extend
the internal symmetry group to the easy-plane SU(3) \cite{DEmidio16},
and explore how the extention of the internal symmetry group affects the multicriticality.
Additionally, as a probe to detect the phase transition
\cite{Quan06,Zanardi06,Zhou08,Yu09,You11,Rossini18,Albuquerque10,Schwandt09},
we resort to the fidelity
\cite{Uhlmann76,Jozsa94,Peres84,Gorin06}
\begin{equation}
\label{fidelity}
	F (H,H+\Delta H) = | \langle H | H+\Delta H\rangle |
,
\end{equation}
where the vectors, $|H\rangle $ and $| H+\Delta H\rangle$,
denote
the ground states with
the proximate interaction parameters, $H$ and $H+\Delta H$, respectively.
The fidelity (\ref{fidelity}) is readily accessible via the exact-diagonalization method,
which yields the ground-state vector $|H \rangle$ explicitly.
According to the elaborated exact-diagonalization study of the two-dimensional 
$XXZ$ and Ising models
\cite{Yu09},
the fidelity-mediated analysis
admits a reliable estimate for the criticality,
although the
available system size $N \le 20$ is rather restricted.

To be specific,
we present the Hamiltonian 
for the two-dimensional easy-plane SU(3) magnet subjected to the transverse field 
\begin{equation}
\label{Hamiltonian}
{\cal H}=
-J\sum_{\langle ij \rangle}
[(1+\eta)
(S^x_iS^x_j+S^y_iS^y_j)+
(1-\eta)
(Q^{zx}_iQ^{zx}_j+Q^{yz}_iQ^{yz}_j)] + 
	H\sum_{i=1}^N Q^{z^2}_i
,
\end{equation}
with the $S=1$-spin operator
${\mathbf S}_i$ placed at each square-lattice point $i=1,2,\dots,N$.
Likewise, the quadrupolar moments at site $i$,
$Q^{x^2-y^2}_i=(S^x_i)^2-(S^y_i)^2$,
$Q^{z^2}_i=\sqrt{3}(S^z_i)^2-\frac{2}{\sqrt3}$,
$Q^{xy}_i=S^x_iS^y_i+S^y_iS^x_i$,
$Q^{yz}_i=S^y_iS^z_i+S^z_iS^y_i$, and
$Q^{zx}_i=S^z_iS^x_i+S^x_iS^z_i$,
are incorporated.
These eight operators constitute the SU$(3)$ algebra
\cite{Lauchli06}
just like the Gell-Mann
matrices.
The summation $\sum_{\langle ij \rangle}$ runs over all possible nearest-neighbor pairs 
$\langle ij \rangle$,
and the coupling constant $J$ sets the unit of energy, $J=1$,
throughout this study. 
The parameters, $\eta$ and $H$,
denote the
anisotropy of the internal symmetry and
the
transverse field, respectively.
The anisotropy $\eta \ne 0$ 
gives rise to the asymmetry between the
$S^{x,y}_i$ and $Q^{yz,zx}_i$ sectors.
Irrespective of the anisotropy $\eta$,
the Hamiltonian commutes with the $z$-axis-rotation generator $\sum_{i=1}^N S^z_i$,
and hence, the Hamiltonian (\ref{Hamiltonian})
describes the easy-plane sector of the SU(3) magnet.

\begin{figure}
\includegraphics[width=13cm]{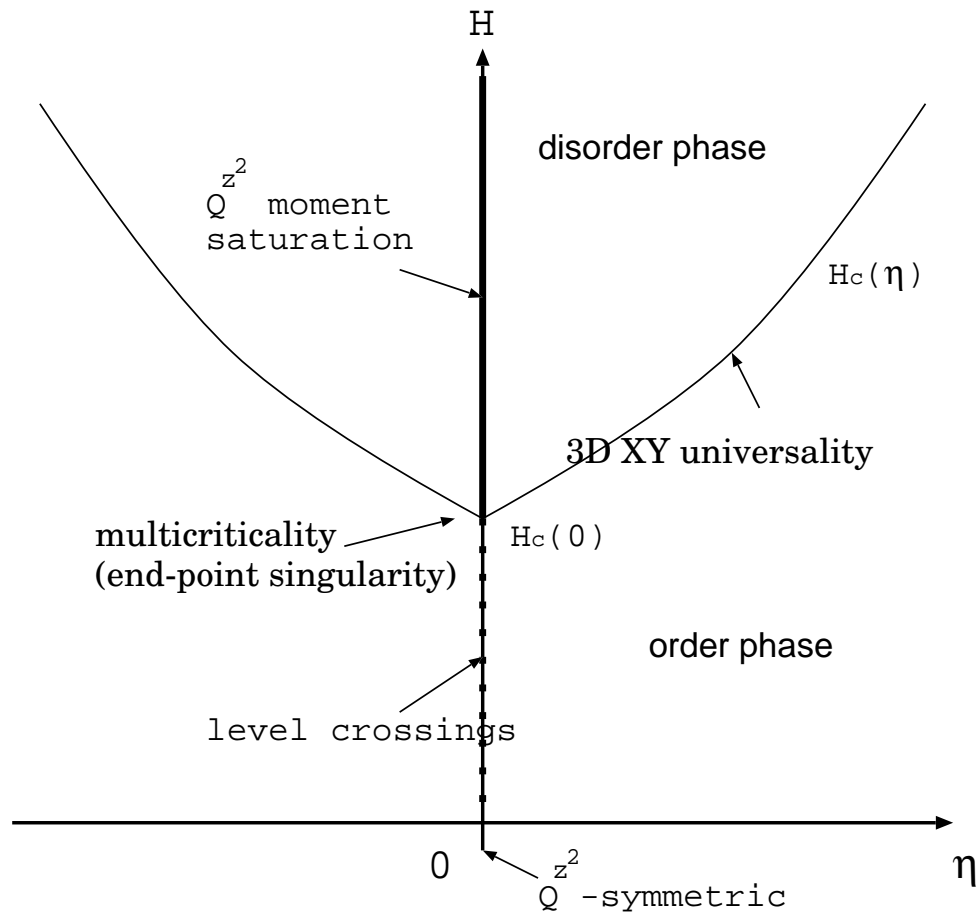}%
\caption{\label{figure1}
A schematic phase diagram of
the easy-plane SU$(3)$
magnet (\ref{Hamiltonian})
for the anisotropy $\eta$
and transverse field $H$ is presented.
The phase boundary $H_c(\eta)$ terminates at the multicritical point 
$H=H_c(\eta=0)=8/\sqrt{3}$  
at $\eta=0$,
and this end-point singularity is our concern.
For small $H<H_c(\eta)$,
the in-plane order, 
$(S^x_i,S^y_i)$
[$(Q^{yz}_i,Q^{zx}_i)$] develops in 
the $\eta>(<)0$ side,
whereas for large $H>H_c(\eta)$, the disorder phase extends.
At the isotropic point $\eta=0$ in between,
	the transverse moment $\sum_{i=1}^N Q^{z^2}_i$ commutes with the Hamiltonian,
and
the ground-state level crossing occurs successively 
\cite{Mukherjee11} up to $H<H_c(0)$.
Above this threshold $H>H_c(0)$,
the magnetization plateau \cite{Zapf14} 
(saturation of 
	the moment $\sum_{i=1}^N Q^{z^2}_i$) 
sets in.
}
\end{figure}

In Fig. \ref{figure1}, 
we present a schematic phase diagram of
the easy-plane SU(3) magnet (\ref{Hamiltonian})
for the anisotropy $\eta$ and the transverse field $H$;
the overall character would resemble
that of the $XY$ model \cite{Henkel84}.
The phase boundary $H_c(\eta)$ separates the
order ($H<H_c(\eta)$)
and disorder ($H>H_c(\eta)$)
phases.
In the order phase,
there appears
the spontaneous magnetization of
the in-plane moment
$(S^x_i,S^y_i)$ [$(Q^{zx}_i,Q^{yz}_i)$]  for the $\eta>(<)0$ regime.
Therefore,
the phase boundary $H=H_c(\eta)$ should belong to the three-dimensional 
(3D) $XY$ universality class 
{\em except} at the multicritical point $\eta=0$.
At $\eta=0$,
the Hamiltonian (\ref{Hamiltonian}) commutes with the transverse moment $\sum_{i=1}^N Q^{z^2}_i$.
Hence, as the transverse field $H$ increases,
the successive level crossings take place as to the ground-state energy
up to $H<H_c(0)(=8/\sqrt{3})$
\cite{Mukherjee11}, and
above the threshold $H>H_c(0)$,
the transverse moment $\sum_{i=1}^N Q^{z^2}_i$ saturates eventually;
see Appendix.
This transition mechanism
is the same as that of the magnetization plateau (saturation of magnetization),
and the power-law singularities have been investigated in depth
\cite{Zapf14}.
Around the threshold $H=H_c(0)$,
the efficiency of the quantum Monte Carlo sampling 
suffers from the slowing-down problem \cite{Kashurnikov99},  
and the aforementioned exact-diagonalization study \cite{Henkel84}
circumvented this difficulty.

Then,
there arises a problem how the phase boundary $H_c(\eta)$
terminates
at
the multicritical point $\eta=0$;
see Fig. \ref{figure2}.
The exact-diagonalization analysis of the $XY$ model
indicates
that
the phase boundary 
$H_c(\eta)$ rises up linearly (monotonically) \cite{Henkel84,Wald15}
around the multicritical point $\eta=0$.
Actually,
the power-law singularity of the phase boundary
\cite{Riedel69,Pfeuty74}
\begin{equation}
\label{crossover_exponent}
	H_c(\eta) -H_c(0)  \sim |\eta|^{1/\phi}
,
\end{equation}
is characterized by the crossover exponent
$\phi \approx 1$ as for the $XY$ magnet \cite{Henkel84,Wald15,Nishiyama19}.
On the one hand,
the large-$N$ theory \cite{Wald15} suggests that the phase boundary 
$H_c(\eta)$ shows
a reentrant behavior.
That is, the phase boundary $H_c(\eta)$ exhibits a non-monotonic
dependence on the anisotropy $\eta$.
So far, only the limiting cases such as the 
$XY$ 
\cite{Henkel84,Wald15,Nishiyama19}
and 
spherical \cite{Wald15}
models have been considered,
and
no information has been provided as to the multicriticality in between.
As the internal symmetry group is enlarged gradually from SU(2),
the phase boundary may become curved convexly,
accompanying with
the suppressed crossover exponent $\phi <1$.
In this paper, 
considering the SU(3) version of the easy-plane magnet,
we investigate the crossover exponent $\phi$ quantitatively
by the agency of
the fidelity $F(H,H+\Delta H)$ (\ref{fidelity}).

The rest of this paper is organized as follows.
In Sec. \ref{section2},
we present the numerical results.
Prior to the detailed analysis of the multicriticality,
we investigate  the case of the fully anisotropic limit $\eta=1$, 
aiming to demonstrate
the performance of our simulation scheme.
In the last section,
we address the summary and discussions.

\section{\label{section2}Numerical results}

In this section, we present the numerical results
for the easy-plane SU$(3)$ magnet subjected to the transverse field (\ref{Hamiltonian}).
We employed
the exact-diagonalization method for the finite-size cluster with
$N \le 5 \times 5$ spins.
The linear dimension of the cluster is given by $L=\sqrt{N}$,
which sets the  fundamental length scale in the subsequent finite-size-scaling analyses.
As a probe to detect the phase transition,
we utilized the fidelity succeptibility
\cite{Quan06,Zanardi06,Zhou08,Yu09,You11,Rossini18,Albuquerque10,Schwandt09}
\begin{equation}
\label{fidelity_susceptibility}
	\chi_F (H)=   -   \frac{1}{L^2} \partial^2_{\Delta H} F(H,H+\Delta H) |_{\Delta H=0}
,
\end{equation}
with the fidelity $F(H,H+\Delta H)$ (\ref{fidelity}).
As was demonstrated 
in Ref.
\cite{Yu09}
for the two-dimensional $XXZ$ and Ising models under the transverse field
with $N \le 20$ spins,
the fidelity susceptibility $\chi_F$ 
 (\ref{fidelity_susceptibility})
admits a reliable estimate of the criticality,
even though the available system size
$N \le 20$ is rather restricted.

In fairness, it has to be mentioned that
the similar scheme was applied to the {\em one-dimensional} $XY$ magnet 
under the transverse field \cite{Mukherjee11}.
In this pioneering study \cite{Mukherjee11}.
the authors took a direct route toward the multicritical point, 
$(\eta,H)\to(0,H_c(0))$.
The $\chi_F$ data
exhibit the intermittent peaks, reflecting the successive
level crossings along the ordinate axis $\eta=0$, as shown in Fig. \ref{figure1}.
In this paper,
to avoid such a finite-size artifact, we took a different route to the multicritical point,
keeping the anisotropy $\eta$ to a finite value.
That is,
based
on the 
the crossover-scaling theory \cite{Riedel69,Pfeuty74},
the anisotropy $\eta$ is scaled properly, as the system size $L$ changes.
As a byproduct, we are able to estimate the crossover exponent $\phi$ quantitatively,
which characterizes the power-law singularity of the phase boundary;
see Fig. \ref{figure2}.
Before commencing detailed crossover-scaling analyses of $\chi_F$,
we devote ourselves to the  fully anisotropic case $\eta=1$
so as to examine the performance of our
simulation scheme.

\subsection{\label{section2_1}
Phase transition point $H_c(1)$ at $\eta=1$:
Fidelity-susceptibility 
analysis
}

In this section,
as a preliminary survey,
setting the anisotropy parameter to the fully anisotropic limit $\eta=1$,
we investigate the order-disorder phase transition
of the easy-plane SU(3) magnet under the transverse field (\ref{Hamiltonian})
via the fidelity susceptibility $\chi_F$ (\ref{fidelity_susceptibility}).
At this point $\eta=1$,
the model (\ref{Hamiltonian}) reduces to 
the spin-$S=1$ $XY$ model with the single-ion anisotropy $D$,
for which a variety of preceding results are available
\cite{Wang05,Pires11,Moura14,Roscilde07,Hammer10,Zhang13}.

In Fig. \ref{figure3},
we present the fidelity susceptibility
$\chi_F$ (\ref{fidelity_susceptibility})
for various values of the transverse field $H$,
and the system sizes, 
($+$) $L=3$,
($\times$) $4$, and
($*$) $5$, with the fixed $\eta=1$.
We see that the fidelity susceptibility exhibits a pronounced signature for
the order-disorder phase transition around $H \approx 6$.

In Fig. \ref{figure4}, we present the approximate
critical point
$H_c^*(L)$
for $1/L^{1/\nu}$ with $\eta=1$ fixed.
Here, the approximate critical point $H_c^*(L)$ 
denotes the location of the fidelity-susceptivity peak
\begin{equation}
\label{approximate_critical_point}
\partial_H
\chi_F(L)|_{H=H_{c}^{*}(L)}=0,
\end{equation}
for each system size $L$.
The power of the abscissa scale $1/\nu$
comes from 
the scaling dimension of the parameter $H$ \cite{Albuquerque10},
and the correlation-length critical exponent $\nu$ is set to
the value of the
3D-$XY$ universality class, $\nu=0.6717$ \cite{Campostrini06,Burovski06};
the validity of this proposition is
examined in the next section.
From Fig. \ref{figure4},
the least-squares fit to these data yields 
an estimate
\begin{equation}
\label{critical_point}
H_c(\eta=1)=6.53(5)
,
\end{equation}
in the thermodynamic limit $L\to\infty$.

This is a good position to address an overview of the related studies.
In Table \ref{table},
we recollect
a number of preceding results for the critical point $H_c(1)$
at the fully anisotropic case $\eta=1$.
As mentioned above, at $\eta=1$,
our model (\ref{Hamiltonian}) reduces to 
the spin-$S=1$ $XY$ model with the single-ion anisotropy $D$,
and the $D$-based results are converted 
via the relation
$H=2D/\sqrt{3}$
so as to match our notation.
As shown in table \ref{table},
the critical point $H_c(1)$ was estimated as
$H_c=6.300$ \cite{Wang05},
$6.628$ \cite{Pires11,Moura14}, and
$6.524(23)$ \cite{Roscilde07}
by means of
the
bosonic mean-field approximation (BMFA),
self-consistent harmonic approximation (SCHA), and
quantum Monte Carlo (QMC) methods, respectively. 
(The BMFA estimate is read off from Fig. 1 of Ref. \cite{Wang05}.)
As a probe to detect the phase transition,
various quantifiers, such as
the 
spontaneous magnetization, 
energy gap, 
and correlation length, 
were
utilized in the respective studies.
As indicated, we also incorporated
the Heisenberg-model \cite{Hammer10,Zhang13} result \cite{Roscilde07},
because ``the anisotropy does not have large effects''
\cite{Wang05} as to the critical point $H_c$.
Our exact-diagonalization (ED) result $H_c=6.53(5) $ 
[Eq.  (\ref{critical_point})]
obtained
via the fidelity susceptibility $\chi_F$
appears to be accordant with
these preceding studies.
Particularly,
our result $H_c=6.53(5)$ [Eq. (\ref{critical_point})]
agrees with
the large-scale-QMC-simulation result $6.524(23)$ \cite{Roscilde07},
validating the $\chi_F$-mediated simulation scheme. 
Encouraged by this finding,
we further explore the critical behavior of $\chi_F$ in the next section.

\subsection{\label{section2_2}
Criticality of the fidelity susceptibility at $\eta=1$}

In this section, we investigate the criticality of the fidelity susceptibility
$\chi_F$ (\ref{fidelity_susceptibility})
at the fully anisotropic limit, $\eta=1$.
To begin with, we set up the finite-size-scaling formula for the fidelity susceptibility
\cite{Albuquerque10}
\begin{equation}
\label{scaling_formula}
\chi_F = L^{\alpha_F/\nu} f \left( (H-H_c)L^{1/\nu} \right)
,
\end{equation}
with a scaling function $f$,
and the fidelity-susceptibility critical exponent $\alpha_F$;
namely, the index $\alpha_F$  describes the singularity $\chi_F \sim |H-H_c|^{-\alpha_F}$
at the critical point $H_c$.
According to Ref. \cite{Albuquerque10},
the critical exponent $\alpha_F$
satisfies the relation
\begin{equation}
\label{scaling_relation}
\alpha_F/\nu = \alpha/\nu+1
,
\end{equation}
with the specific-heat critical index $\alpha$;
namely, the index $\alpha$
describes the singularity of the specific heat as $C \sim |H-H_c|^{-\alpha}$.
We postulate that the the criticality belongs to the
3D-$XY$
universality class \cite{Campostrini06,Burovski06} for the anisotropic regime $\eta \ne 0 $.
Putting the existing values \cite{Campostrini06},
$\alpha=0.0151$ and $\nu=0.6717$,
for the
3D-$XY$ universality class into the scaling relation (\ref{scaling_relation}),
we arrive at
\begin{equation}
\label{critical_exponent}
\alpha_F/\nu=0.9775    
.
\end{equation}
Notably, this index $\alpha_F/\nu=0.9775$ is larger than
that of the specific heat, $\alpha / \nu=-0.0225$ \cite{Campostrini06};
actually, the latter takes a negative value.
The scaling parameters,
$\alpha_F/\nu$ and $\nu$, appearing
in the expression (\ref{scaling_formula})
are all fixed,
and we are able to carry out the scaling analysis of $\chi_F$ unambiguously.

In Fig. \ref{figure5}
we present the scaling plot,
$(H-H_c)L^{1/\nu}$-$\chi_F L^{-\alpha_F/\nu}$,
for various $H$
and system sizes, 
($+$) $L=3$,
($\times$) $4$, and
($*$) $5$,
with the fixed $\eta=1$.
Here, the scaling parameters are set to
$H_c=6.53$ [Eq. (\ref{critical_point})],
$\nu=0.6717$ \cite{Campostrini06}, and
$\alpha_F/\nu=0.9775$ [Eq. (\ref{critical_exponent})].
The scaled data appear to collapse into the scaling curve satisfactorily,
confirming that the simulation data already enter into the scaling regime.
We stress that no {\it ad hoc} parameter adjustment is undertaken in the present
scaling analysis.

We address
a number of remarks.
First,
the scaling plot, Fig. \ref{figure5},
indicates that the phase transition belongs to the
3D-$XY$ universality class.
So far, 
as for the {\it Heisenberg antiferromagnet} with the single-ion anisotropy $D$,
it has been claimed that
the $D$-driven singularity belongs to the
3D-$XY$ universality class \cite{Roscilde07,Zhang13}.
Our simulation result shows that
the quantum $XY$ ferromagnet
is also under the reign of the 3D-$XY$
universality class.
Second,
from the scaling plot, Fig. \ref{figure5},
we see that
corrections to finite-size scaling are rather suppressed.
Actually,
as first noted by Ref. \cite{Yu09},
the fidelity susceptibility detects the underlying singularity
clearly out of the subdominant contributions.
Last, 
a key ingredient is that
$\chi_F$'s singularity $\alpha_F/\nu=0.9775$ 
[Eq. (\ref{critical_exponent})]
is substantially larger
than that of the specific heat, $\alpha/\nu =-0.0225 $ \cite{Campostrini06};
actually,
the latter 
takes a negative value as for the 3D-$XY$ universality.
In this sense, the former
is appropriate as a quantifier for the criticality.

\subsection{\label{section2_3}
Crossover-scaling plot of the fidelity susceptibility 
around $\eta=0$: Analysis of the crossover exponent $\phi$}

We then turn to the analysis of the multicriticality at $\eta=0$.
For that purpose,
we extend the above scaling formalism (\ref{scaling_formula})
to the crossover-scaling formula \cite{Riedel69,Pfeuty74}
\begin{equation}
\label{crossover_scaling_formula}
\chi_F = L^{\dot{\alpha}_F/\dot{\nu}}
   g \left(
(H-H_c(\eta))L^{1/\dot{\nu}} ,
\eta L^{\phi/\dot{\nu}} 
   \right)
,
\end{equation}
with 
yet another controllable parameter $\eta$, the
accompanying crossover exponent $\phi$,
and a scaling function $g$.
Here, the indices, $\dot{\alpha}_F$
and $\dot{\nu}$, denote
the fidelity-susceptibility and correlation-length critical exponents, respectively, 
right at the multicritical point $\eta=0$.
The meaning of the
second argument of the crossover-scaling formula (\ref{crossover_scaling_formula}),
$\eta L^{\phi/\dot{\nu}}$, is as follows.
The correlation-length critical exponent $\dot{\nu}$ describes
the singularity of the length scale $L$ 
as $L \sim |H-H_c(0)|^{-\dot{\nu}}$. This relation leads to
the physically convincing expression 
$\eta L^{\phi/\dot{\nu}} \sim \eta/|H-H_c(0)|^\phi$.
Now, it is apparent that the crossover exponent $\phi$ describes the
mutual relationship between $H$ and the new entity $\eta$.

Before carrying out the crossover-scaling analyses,
we fix the values of the critical indices appearing in the formula (\ref{crossover_scaling_formula}).
As mentioned in Introduction, the phase transition along the ordinate axis $\eta=0$
is essentially the same as that of the magnetization plateau,
and the power-law singularities have been investigated in considerable detail \cite{Zapf14}
as follows.
The corelation-length critical exponent was determined as $\dot{\nu}=1/2$ \cite{Zapf14,Adamski15,Hoeger85}.
This index immediately yields $\dot{\alpha}_F/\dot{\nu}=3$ via the scaling relation 
$\dot{\alpha}_F/\dot{\nu}=\dot{\alpha}/\dot{\nu}+z$ \cite{Albuquerque10}.
Here, the dynamical critical exponent takes the value $z=2$ \cite{Zapf14},
and the specific-heat critical exponent 
$\dot{\alpha}=1/2$ comes from the idea that
the field-induced magnetization, $\sim \sqrt{H-H_c}$ \cite{Zapf14}, is regarded as the internal energy.
(Note that the first derivative of the ground-state energy corresponds to the
``internal energy'' in the classical statistical mechanical context.)
The above consideration now completes the prerequisite for the crossover-scaling analysis.
The remaining index $\phi$ is to be adjusted 
so as to achieve an alignment of the crossover-scaled $\chi_F$ data.

In Fig. \ref{figure6},
we present the crossover-scaling plot,
$(H-H_c(\eta))L^{1/\dot{\nu}}$-$\chi_F L^{-\dot{\alpha}_F/\dot{\nu}}$, 
for various system sizes,
($+$) $L=3$,
($\times$) $4$, and
($*$) $5$,
with $\dot{\nu}=1/2$ and $\dot{\alpha}_F/\dot{\nu}=3$ determined above.
Here, we fixed the second argument of the scaling formula 
(\ref{crossover_scaling_formula})
to a constant value $\eta L^{\phi /\dot{\nu} }=5$
with an optimal crossover exponent $\phi=0.8$,
and the parameter $H_c(\eta)$ was determined via the same scheme
as that of Sec. \ref{section2_1}.
From Fig. \ref{figure6},
we see that the
crossover-scaled data fall into the scaling curve satisfactorily.
Particularly, the $L=4$ ($\times$) and
$5$ ($*$) data are about to overlap each other.
Such a feature suggests that
the choice $\phi=0.8$ is a feasible one.

Likewise, setting the crossover exponent
to $\phi=0.7$,
in Fig. \ref{figure7},
we present
the crossover-scaling plot,
$(H-H_c(\eta))L^{2}$-$\chi_F L^{-3}$, 
for various system sizes $L=3,4,5$;
the symbols are the same as those 
of Fig. \ref{figure6}.
Here, the second argument of the crossover-scaling
formula (\ref{crossover_scaling_formula})
is fixed to $\eta L^{2\phi}=3.62$ with a proposition $\phi=0.7$.
For such a small value of $\phi=0.7$, the crossover-scaled data get scattered;
particularly, the left-side slope becomes dispersed,
as compared to that of Fig. \ref{figure6}.
Similarly,
under the setting $\phi=0.9$,
in Fig. \ref{figure8},
we
present the crossover-scaling plot,
$(H-H_c(\eta))L^{2}$-$\chi_F L^{-3}$, 
for various system sizes $L=3,4,5$; 
the symbols are the same
as those of Fig. \ref{figure6}.
Here,
the second argument of the crossover-scaling formula
is set to a constant value
$\eta L^{2\phi}=6.9$ with $\phi = 0.9$.
For such a large value of $\phi=0.9$,
the crossover-scaled data become scattered;
actually,
the hill-top data split up.
Considering that the above cases, Figs. \ref{figure7}
and \ref{figure8}, yield the lower and upper bounds, respectively,
for the estimate of $\phi$,
we
conclude that the crossover exponent lies within 
\begin{equation}
\label{crossover_exponent2}
\phi=0.8(1)
.
\end{equation}
Our result excludes the possibility that the 
phase boundary $H_c(\eta)$
rises up linearly $\phi \approx 1$,
as observed for the $XY$ (namely, easy-plane SU(2)) magnet \cite{Henkel84,Wald15,Nishiyama19}.
Rather, 
as for the extended internal symmetry group,
the phase boundary is curved convexly,
accompanying with
a slightly suppressed
crossover exponent (\ref{crossover_exponent2}).

A number of remarks are in order.
First,
the underlying physics behind the crossover-scaling plot,
Fig. \ref{figure6},
differs from that of the fixed-$\eta$ scaling, Fig. \ref{figure5}.
Actually, the former scaling dimension
$\dot{\alpha}_F/\dot{\nu}=3$ 
is 
much larger than that of the latter,
$\alpha_F/\nu=0.9775$ [Eq. (\ref{critical_exponent})].
Therefore,
the index $\phi$ has to be adjusted carefully, and
the collapse of the crossover-scaled data points, Fig. \ref{figure6}, 
is by no means coincidental.
In this sense,
the present crossover-scaling analysis captures the characteristics of
the multicritical behavior.
Second,
because the fidelity-susceptibility approach does not rely
on any presumptions as to the order parameter involved,
it detects both singularities, $\dot{\alpha}/\dot{\nu}=3$
and $\alpha/\nu=0.9775$ [Eq. (\ref{critical_exponent})], in a unified manner.
Moreover, as mentioned in Sec. \ref{section2_2},
the fidelity susceptibility is less affected by
corrections to scaling \cite{Yu09},
and it 
picks up the underlying singularity out of the subdominant contributions.
Last,
as explained in Introduction, so far,
only the extremum cases, namely, the large-$N$ \cite{Wald15}
and easy-plane-SU(2) \cite{Henkel84,Wald15,Nishiyama19} symmetry 
groups, have been considered,
and the multicritical behavior in between remains unclear.
Our result $\phi=0.8(1)(<1)$ [Eq. (\ref{crossover_exponent2})]
indicates that the phase boundary $H_c(\eta)$ 
is curved
convexly, as the internal symmetry group is extended to the easy-plane SU(3). 
Because the reentrant behavior occurs only in
$1<d \le 2.065\dots$ dimensions even for the large-$N$ case \cite{Wald15},
it is reasonable that the {\em slightly} enlarged SU(3) group 
does not lead to the reentrant behavior immediately in $d=2$ dimensions.

\section{\label{section3}Summary and discussions}

\begin{figure}
\includegraphics[width=13cm]{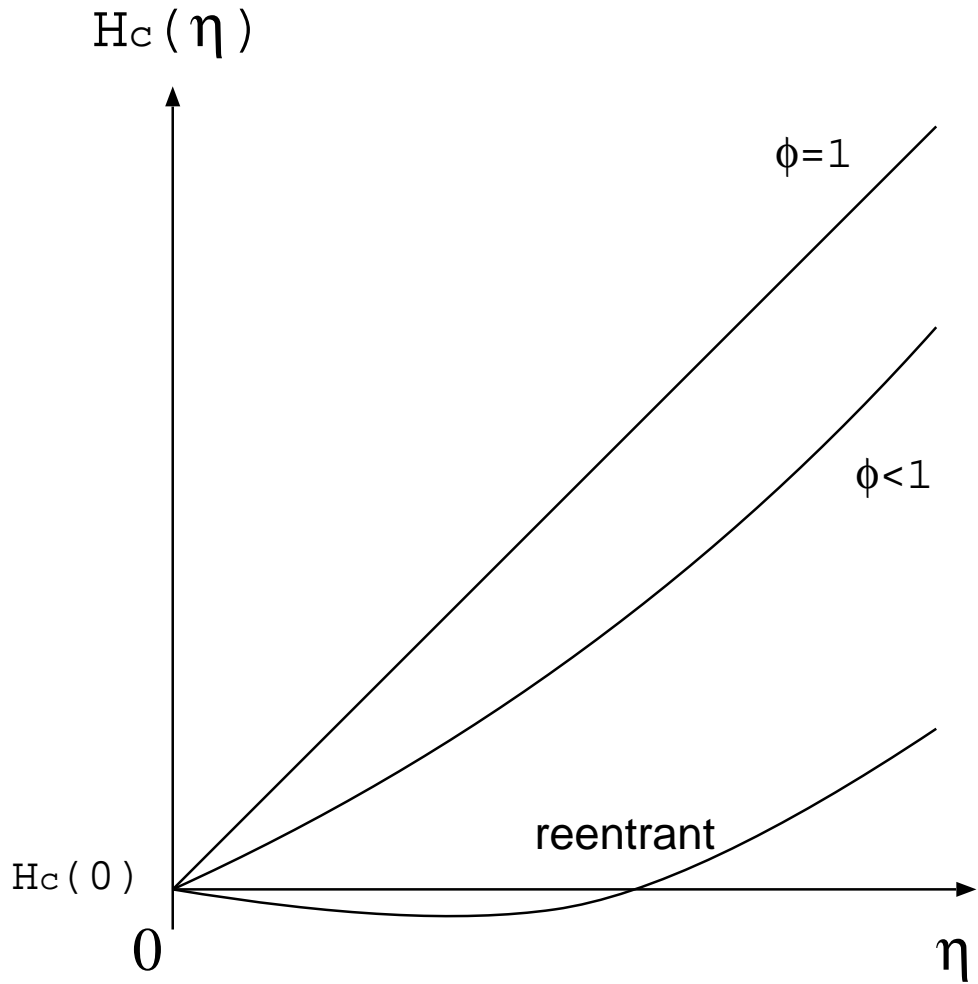}%
\caption{\label{figure2}
The multicriticality (end-point singularity) 
of the phase boundary $H_c(\eta)$
at $\eta=0$ is characterized 
by the crossover exponent $\phi$ 
\cite{Riedel69,Pfeuty74} such as 
	$H_c(\eta) -H_c(0) \sim |\eta|^{1/\phi} $ (\ref{crossover_exponent}).
As for the $XY$ model (namely, the easy-plane SU$(2)$ magnet),
the exact-diagonalization simulation 
suggests that the phase boundary rises up
linearly with $\phi \approx 1$ \cite{Henkel84,Wald15,Nishiyama19}.
On the one hand,
the large-$N$ analysis admits the reentrant (non-monotonic)
behavior \cite{Wald15}.
As indicated,
the reentrant behavior leads to  a counterintuitive picture 
such that the disorder phase is induced by the lower internal symmetry group
$\eta\ne 0$
around the multicritical point.
It is anticipated that for the SU$(3)$ case,
the phase boundary is curved convexly with a slightly suppressed 
crossover exponent $\phi<1$.
}
\end{figure}

\begin{figure}
\includegraphics[width=13cm]{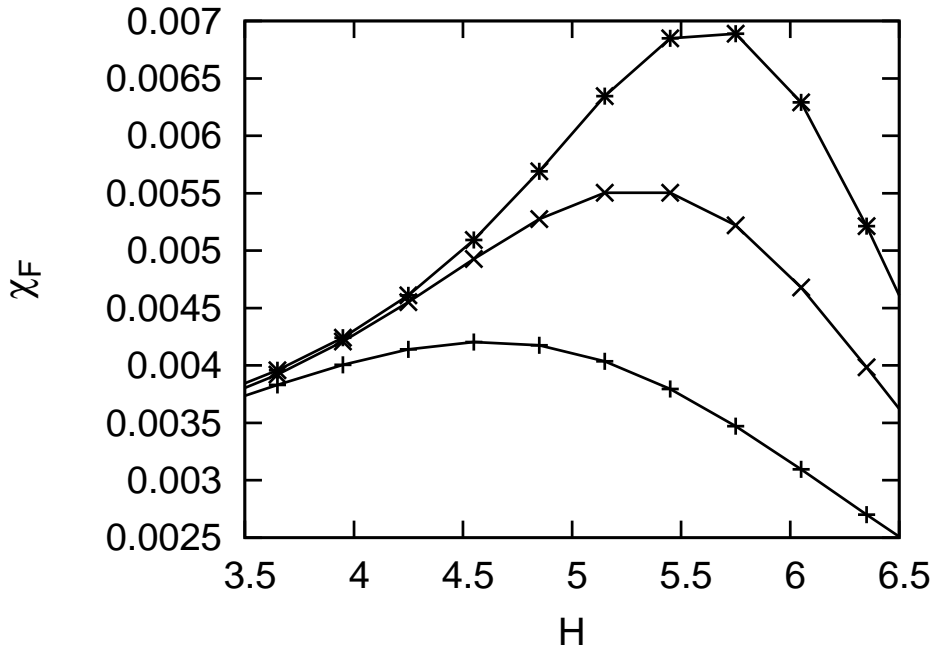}%
\caption{\label{figure3}
The fidelity susceptibility $\chi_F$
(\ref{fidelity_susceptibility})
is plotted for various values of the transverse field $H$ and the system sizes, 
($+$) $L=3$,
($\times$) $4$, and
($*$) $5$,
with the fixed anisotropy parameter $\eta=1$.
The fidelity susceptibility indicates a notable peak around
the critical point $H \approx 6$.
}
\end{figure}

\begin{figure}
\includegraphics[width=13cm]{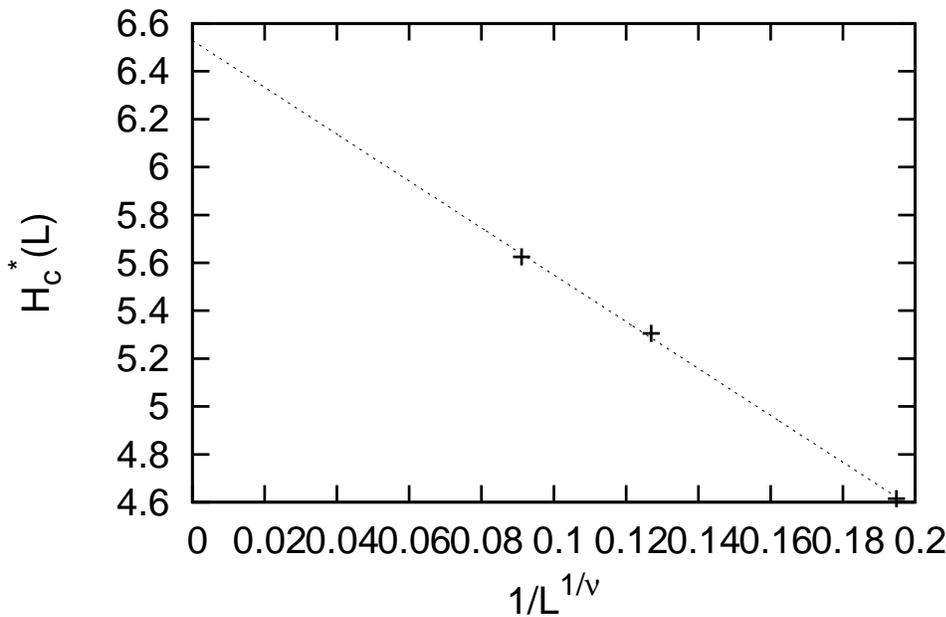}%
\caption{\label{figure4}
The approximate critical point 
$H_{c}^*(L)$
(\ref{approximate_critical_point})
is plotted for $1/L^{1/\nu}$ with the fixed anisotropy parameter $\eta=1$.
Here, the correlation-length critical exponent is set to 
the 3D-$XY$-universality value $\nu=0.6717$ \cite{Campostrini06,Burovski06}.
The least-squares fit to these data yields
an estimate $H_c=6.53(5)$ in the thermodynamic limit $L \to \infty$.
}
\end{figure}

\begin{figure}
\includegraphics[width=13cm]{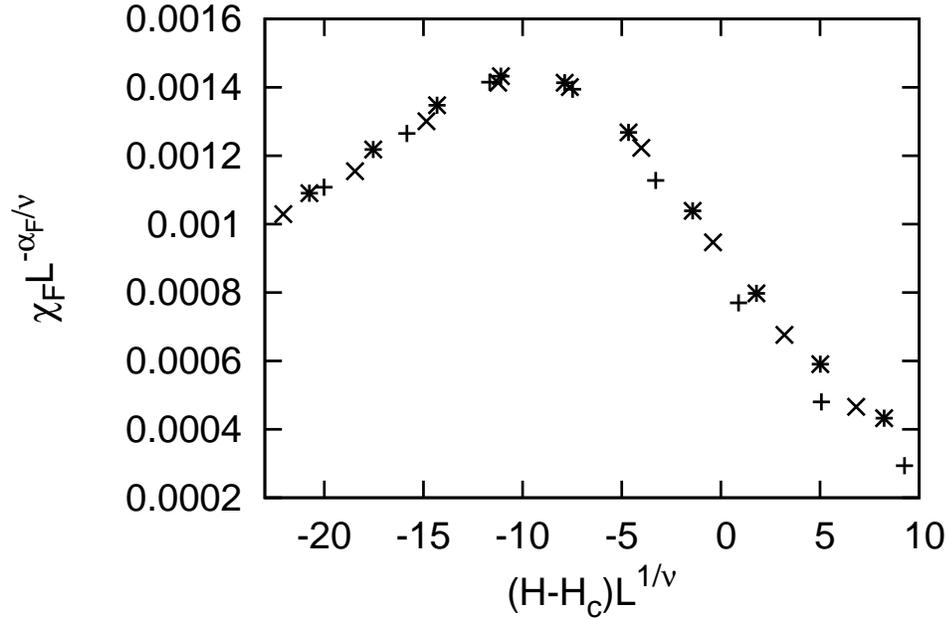}%
\caption{\label{figure5}
The scaling plot,
$(H-H_c)L^{1/\nu}$-$\chi_F L^{- \alpha_F/\nu}$,
is presented 
for various system sizes,
($+$) $L=3$,
($\times$) $4$, and
($*$) $5$,
with the fixed $\eta=1$;
see the scaling formula
(\ref{scaling_formula}).
Here, the scaling parameters 
are set to
	$H_c=6.53$ [Eq. (\ref{critical_point})], 
$\nu=0.6717$ \cite{Campostrini06,Burovski06}
and 
	$\alpha_F/\nu=0.9775$ [Eq. (\ref{critical_exponent})].
The scaled data collapse into the scaling curve satisfactorily, suggesting that
the criticality belongs to the 3D-$XY$ universality class.
}
\end{figure}

\begin{figure}
\includegraphics[width=13cm]{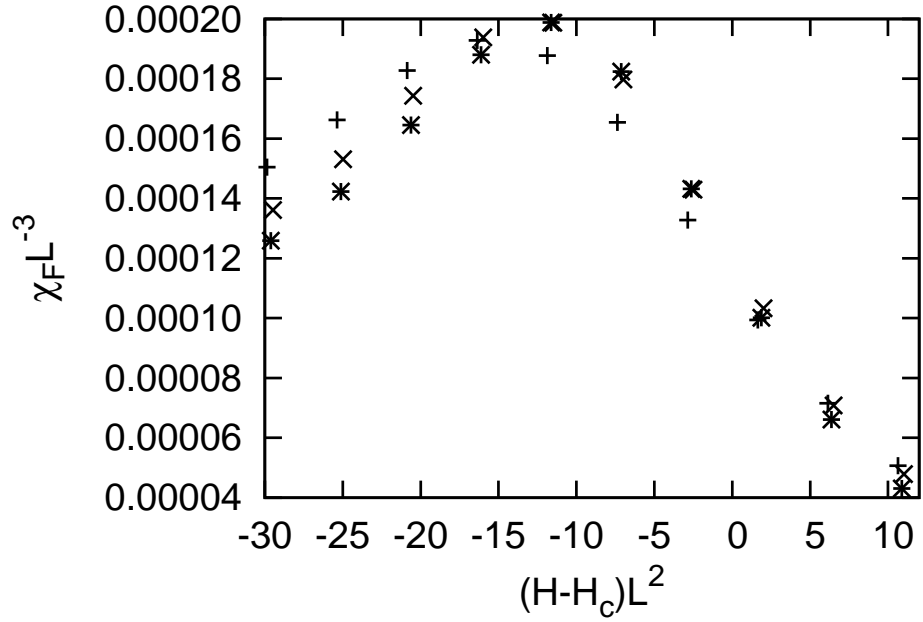}%
\caption{\label{figure6}
The crossover-scaling plot,
$(H-H_c(\eta))L^{1/\dot{\nu}}$-$\chi_F L^{- \dot{\alpha_F}/\dot{\nu}}$,
is presented 
for various system sizes,
($+$) $L=3$,
($\times$) $4$, and
($*$) $5$,
with 
$\dot{\nu}=1/2$ \cite{Zapf14}
and 
$\dot{\alpha}_F/\dot{\nu}=3$; see text for details.
Here, the second argument of 
the crossover-scaling formula
(\ref{crossover_scaling_formula})
is fixed to $\eta L^{\phi/\dot{\nu}}=5$
with the crossover exponent $\phi=0.8$.
The crossover-scaled data fall into the scaling curve satisfactorily;
	particularly, the $L=4$ ($\times$)
	and $5$ ($*$) data are about to overlap each other
	under the setting, $\phi=0.8$. 
}
\end{figure}

\begin{figure}
\includegraphics[width=13cm]{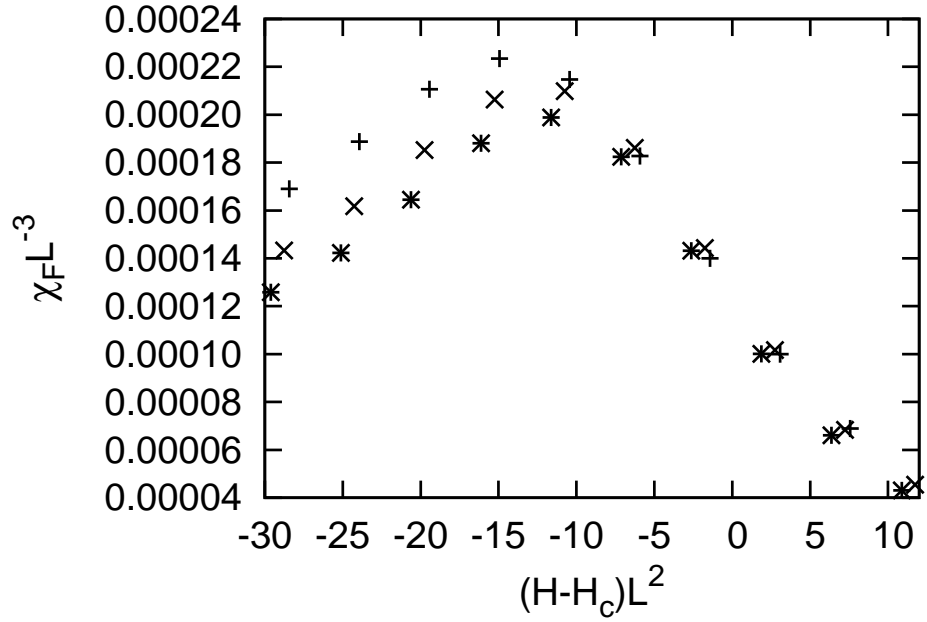}%
\caption{\label{figure7}
The crossover-scaling plot,
$(H-H_c(\eta))L^{1/\dot{\nu}}$-$\chi_F L^{- \dot{\alpha_F}/\dot{\nu}}$,
is presented 
for various system sizes,
($+$) $L=3$,
($\times$) $4$, and
($*$) $5$,
	with $\dot{\nu}=1/2$ and $\dot{\alpha}_F/\dot{\nu}=3$.
Here, the second argument of 
the crossover-scaling formula
(\ref{crossover_scaling_formula})
is fixed to $\eta L^{\phi/\dot{\nu}}=3.62$
with the crossover exponent $\phi=0.7$.
The left-side slope gets scattered under
such a small value of $\phi=0.7$.
}
\end{figure}

\begin{figure}
\includegraphics[width=13cm]{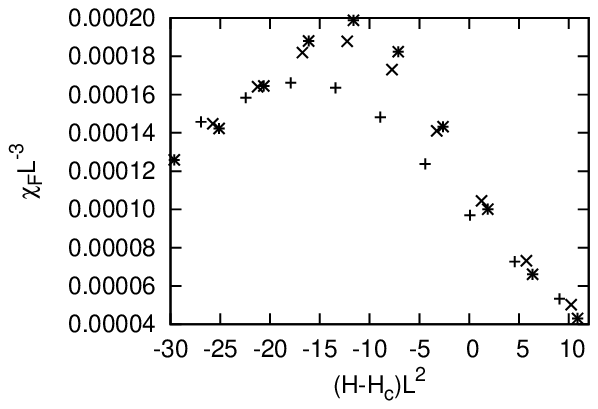}%
\caption{\label{figure8}
The crossover-scaling plot,
$(H-H_c(\eta))L^{1/\dot{\nu}}$-$ \chi_ F L^{ - \dot{\alpha_F}/\dot{\nu}}$,
is presented 
for various system sizes,
($+$) $L=3$,
($\times$) $4$, and
($*$) $5$,
	with $\dot{\nu}=1/2$ and $\dot{\alpha}_F/\dot{\nu}=3$.
Here, the second argument of 
the crossover-scaling formula
(\ref{crossover_scaling_formula})
is fixed to $\eta L^{\phi/\dot{\nu}}=6.9$
with the crossover exponent $\phi=0.9$.
The hill-top data split up for such a large 
value of 
$\phi=0.9$.
}
\end{figure}

The two-dimensional easy-plane SU(3) magnet under the transverse field
(\ref{Hamiltonian})
was investigated with the exact-diagonalization method.
Because the method allows direct access to the ground state,
we do not have to care about the
anisotropy between the real-space and imaginary-time directions 
rendered by the
dynamical critical exponent $z=2(\ne 1)$ \cite{Zapf14} at the multicritical point $\eta=0$.
Our main concern is to
reconcile the discrepancy between the
large-$N$ theory 
\cite{Wald15}
and the exact-diagonalization analysis of the $XY$ magnet
\cite{Henkel84,Wald15,Nishiyama19}
as to the multicriticality (Fig. \ref{figure2}).
For that purpose, we consider the SU(3) version of the easy-plane magnet
(\ref{Hamiltonian}),
and performed the exact-diagonalization simulation 
by the agency of
the fidelity susceptibility $\chi_F$ (\ref{fidelity_susceptibility}).
The fidelity susceptibility has an advantage in that it detects the phase transition 
sensitively, as compared to that of the specific heat \cite{Yu09,Albuquerque10}.
As a preliminary survey, setting the anisotropy parameter
to $\eta=1$,
we estimate the critical point via $\chi_F$ 
as
$H_c(1)=6.53(5)$ [Eq. (\ref{critical_point})].
This result is comparable to the preceding results,
$6.300$ \cite{Wang05},
$6.628$ \cite{Pires11,Moura14}, and
$6.524(23)$ \cite{Roscilde07}, estimated with the 
BMFA,
SCHA, and
QMC methods, respectively.
Particularly, our result $H_c(1)=6.53(5)$
agrees with the large-scale-QMC-simulation result $6.524(23)$ \cite{Roscilde07},
validating the $\chi_F$-mediated simulation scheme.
Thereby, 
we cast the $\chi_F$ data into the crossover-scaling formula
(\ref{crossover_scaling_formula}) with the anisotropy $\eta$ scaled properly.
Adjusting the crossover exponent $\phi$ carefully, 
we attain an alignment of the crossover-scaled data points
for
$\phi=0.8(1)$ [Eq. (\ref{crossover_exponent2})].
This result $\phi < 1$
indicates that the phase boundary $H_c(\eta)$ gets curved convexly 
around the multicritical point $\eta=0$,
as the internal symmetry group is extended to SU(3).
Actually, only in $1< d \le 2.065 \dots$ dimensions,
the reentrant behavior is realized
even for the large-$N$ case \cite{Wald15}.
Hence,
it is reasonable that the slightly enlarged internal symmetry
does not lead to the reentrant behavior immediately.

We
conjecture that 
the phase boundary may exhibit a quadratic curvature,
$\phi=0.5$, eventually
for an extremely large internal symmetry group,
and above this threshold,
the reentrant behavior should set in.
It is thus tempting to consider the fractional-dimensional
($1< d <2$) system realized effectively
by the power-law-decaying interactions \cite{Defenu17}.
In such a fractional-dimensional system,
the reentrant behavior may come out even for the $XY$ model.
This problem is left for the future study.

\ack

This work was supported by a Grant-in-Aid
for Scientific Research (C)
from Japan Society for the Promotion of Science
(Grant No. 20K03767).

\begin{table}
\caption{\label{table}
A number of
preceding results for the critical point $H_c(1)$ at $\eta=1$
are recollected.
At $\eta=1$,
the model (\ref{Hamiltonian}) reduces to the spin-$S=1$
$XY$ model with the single-ion anisotropy $D$.
These $D$-based results are converted
so as to match our notation
via the relation $H=2D/\sqrt{3}$.
So far, a variety of technieus,
such as the 
bosonic mean-field approximation (BMFA) \cite{Wang05},
self-consistent harmonic approximation (SCHA) \cite{Pires11,Moura14},
and
quantum Monte Carlo (QMC) \cite{Roscilde07} methods,
have been employed, and these results are comparable to 
the present
exact diagonalization (ED) result.
(The BMFA estimate is read off from Fig. 1 of Ref. \cite{Wang05}.)
In the respective studies, as a probe to detect the phase transition,
various quantifiers, such as the 
	spontaneous magnetization,
	energy gap, 
correlation length and fidelity susceptibility,
were utilized.
Because ``the anisotropy does not have large effects''
\cite{Wang05},
the large-scale QMC result
 \cite{Roscilde07}
for the Heisenberg model 
\cite{Hammer10,Zhang13}
is shown as well.
}
\begin{tabular}{@{}llll}
\br
Method & Quantifier & Model & $H_c(\eta=1)$ \\
\mr
BMFA \cite{Wang05} & spontaneous magnetization & $XY$ & $6.300$ \\
SCHA \cite{Pires11,Moura14} & energy gap & $XY$ & $6.628$ \\
QMC \cite{Roscilde07} & correlation length & Heisenberg & $6.524(23)$ \\
	ED (this work) & fidelity susceptibility &  $XY$ & $6.53(5)$ \\
\br
\end{tabular}

\end{table}

\appendix

\section{\label{appendix}
Transition point $H_c=8/\sqrt{3}$ at $\eta=0$}

At the isotropic point $\eta=0$,
the location of the multicritical point $H_c(0)=8/\sqrt{3}$ 
is obtained analytically as follows.
For sufficiently large $H$,
the ground state is given by 
the direct product
$\otimes_i^N |0\rangle_i$
of the local base $ |0\rangle_i $,
which
satisfies $S^z_i|m\rangle_i=m|m\rangle_i$
($m=-1,0,1$) at each site $i$.
Due to the strict selection rule at $\eta=0$,
magnons' pair creation is prohibited.
Thus,
the single magnon at site $j$,
$|j\rangle =|1 \rangle_j \otimes (\otimes_{i \ne  j}^N |0\rangle_i)$,
propagates coherently
through the transfer amplitude $-2J$ over the nearest neighbors,
obeying the
dispersion relation 
$-4J(\cos k_x + \cos k_y)+\sqrt{3}H$ (${\mathbf k}$: wave number)
above the ground state.
Therefore, at $H_c(\eta=0)=8/\sqrt{3}$, 
the band gap closes in a way reminiscent of the metal-insulator transition.
This transition mechanism is precisely the same as that of the
magnetization plateau \cite{Zapf14}
(saturation of the magnetization).
A notable point is that
the
dynamical critical exponent takes $z=2$,
because the {\em quadratic} band bottom touches the ground-state energy level.
In other words,
the symmetry between the real-space and imaginary-time directions is violated
by $z \ne 1$.
It is a benefit of
the exact-diagonalization method 
that the method allows direct access to the
ground state,
for which
the imaginary-time system size is infinite.


\end{document}